\documentclass[%
 aip,
 amsmath,amssymb,
 reprint,%
]{revtex4-2}

\usepackage{times}
\usepackage{multirow}
\usepackage{siunitx}
\usepackage{caption}

\usepackage{graphicx}
\usepackage{dcolumn}
\usepackage{bm}

\usepackage[utf8]{inputenc}
\usepackage[T1]{fontenc}
\usepackage{mathptmx}

\begin{document}

\preprint{AIP/123-QED}

\title{Ultracompact energy transfer in anapole-based metachains}
\author{T. C. Huang}
\author{B. X. Wang}
\author{W. B. Zhang}
\author{C. Y. Zhao}
\email{changying.zhao@sjtu.edu.cn}

\affiliation{Institute of Engineering Thermophysics, School of Mechanical Engineering and Key Laboratory of Power Machinery and Engineering of Ministry of Education, Shanghai Jiao Tong University, Shanghai 200240, China}

\date{\today}

\begin{abstract}
Realization of electromagnetic energy confinement beyond the diffraction limit is of paramount importance for novel applications like nano-imaging, information processing, and energy harvest. Current approaches based on surface plasmon polaritons and photonic crystals are either intrinsically lossy or with low coupling efficiency. Herein, we successfully address these challenges by constructing an array of nonradiative anapoles that originate from the destructive far-field interference of electric and toroidal dipole modes. The proposed metachain can achieve ultracompact (1/13 of incident wavelength) and high-efficiency electromagnetic energy transfer without the coupler. We experimentally investigate the proposed metachain at mid-infrared and give the first near-field experimental evidence of anapole-based energy transfer, in which the spatial profile of anapole mode is also unambiguously identified at nanoscale. We further demonstrate the metachain is intrinsically lossless and scalable at infrared wavelengths, realizing a 90$^\circ$ bending loss down to 0.32 dB at the optical communication wavelength. The present scheme bridges the gap between the energy confinement and transfer of anapoles, and opens a new gate for more compactly integrated photonic and energy devices, which can operate in a broad spectral range. 
\end{abstract}

\maketitle

\section*{Introduction}

Overcoming the optical diffraction limit and realizing high-efficiency electromagnetic (EM) energy transfer at the subwavelength scale has attracted much interest due to the broad applications in energy conversion\cite{PhysRevLett.97.157403,koenderink2015nanophotonics}, integrated optics\cite{cheben2018subwavelength,konoike2016demand,chen2019microsphere}, optical modulation\cite{zhang2018generating}, biosensing\cite{rodrigo2015mid}, and so on. To avoid diffraction losses and minimize parasitic interactions between different elements\cite{Rasskazov:14}, a compact device for energy transfer is expected to feature strong EM fields confinement and low intrinsic/radiative loss simultaneously. Plasmonic structures like metallic nanowires\cite{oulton2008hybrid}, nanoholes\cite{Zhang:10,PhysRevB.91.235406}, nanogrooves\cite{doi:10.1063/1.1953877,kildishev2013planar}, and nano strips\cite{doi:10.1063/1.1542944,1589080}, which support strongly confined surface plasmon polaritons (SPPs) at tenth to hundredth of the operating wavelength. However, plasmonic structures notoriously suffer from the intrinsic ohmic loss\cite{palik1998handbook}. The dispersive permittivity of metallic material makes these devices difficult to be tuned to different operating wavelengths solely by their geometry parameters. As an alternative method to transfer energy compactly, dielectric photonic crystals (PCs) require high-precision fabrication, since they heavily depend on deliberate arrangements of unit cells to realize strong energy confinement by guided modes\cite{russell2003photonic}, coupled-cavity modes\cite{doi:10.1063/1.2345374}, topological edge states\cite{barik2018topological}, etc. Furthermore, PC waveguides usually require an extra coupler or converter to enhance out-of-plane coupling efficiency \cite{politi2008silica}, which severely increases the complexity of photonic devices.

As an analog of static anapoles introduced by Zeldovich \cite{zel1958electromagnetic} in the context of parity violation in nuclear physics, optical anapoles feature strong near-field energy confinement and suppressed radiative loss due to the destructive far-field interference of simultaneously excited toroidal dipole (TD) and electric dipole (ED) modes \cite{miroshnichenko2015nonradiating,kaelberer2010toroidal}. A fascinating feature is that anapole can be excited in isolated dielectric particles by normally incident plane wave, which avoids intrinsic loss of SPPs and deliberate lattice arrangements or the requirement of coupler in PCs. As a result, low-loss anapoles with profound near-field enhancement provide an excellent platform for realizing various intriguing physical phenomena and applications, like enhanced nonlinear effect\cite{doi:10.1021/acsnano.6b07568,zhang2020anapole,xu2018boosting}, cloaking\cite{doi:10.1002/lpor.201800005}, high-Q supercavities\cite{PhysRevLett.119.243901,doi:10.1063/1.5078576}, and nanolasers\cite{gongora2017anapole}. On the other hand, the near-field coupling of anapoles is expected to transfer energy in the long-range\cite{mazzone2017near}, offering a promising alternative for current waveguides based on lossy SPPs and complex PCs. However, most current studies focus on the excitation of anapole state in different structures\cite{miroshnichenko2015nonradiating,cui2020anapole,kim2015radiative}, when the theoretical and experimental demonstrations of near-field anapoles coupling and resulted energy transfer remain elusive.

To fill this gap, in this work, we theoretically and experimentally demonstrate that anapole-based metachains consisting of subwavelength silicon disks with air gaps can compactly transfer EM energy in the long range with high efficiency. We find that an air gap structure can significantly enhance the energy confinement inside the isolated disks, which further improves the efficiency of the metachain. Meanwhile, we experimentally evaluated the energy transfer efficiency spectrum of the proposed metachain by the scattering-type scanning near-field microscopy (s-SNOM) at mid-infrared frequencies. The near-field signals manifest the anapole frequencies and depict the distributions of excited modes, from which we experimentally confirm the relationship between the anapole excitation and the high-efficiency energy transfer. Moreover, the proposed metachain is scalable\cite{walia2015flexible} to work at communication wavelength with low loss even when it is designed as a sharp bend. Normally incident plane waves can directly couple with the subwavelength silicon disks, eliminating the need for an extra coupler. This work provides the first near-field experimental evidence of EM energy transmission due to the coupling of anapoles, which opens a novel pathway for realizing ultracompact photonic communications and energy management, such as photonic network-on-chip, quantum memory for entangled photons, chip-scale radiative cooling, etc. 
 
\section*{Excitation of anapoles in isolated meta-atoms}

Since the energy confinement depends on the anapole states excited in each dielectric particle, the effect of geometrical parameters on the excitation of anapoles should be first addressed. The structures of proposed metachains consisting of periodic silicon particles are illustrated in Fig. 1, when the height of particles is 1/5 of the radius ($h$= $R$/5). The refractive index of silicon can be approximately regarded as a constant 3.5 at visible-infrared frequencies\cite{palik1998handbook}, which makes proposed metachains scalable according to the size parameters. Thus, we can use the normalized size and frequency to discuss the performance of different particles, which provides a guideline to design the metachains of interest.

By introducing an air slot structure in a thin dielectric disk, one can excite anapole states\cite{doi:10.1021/acs.nanolett.7b04200} with higher $Q$-factors\cite{gladyshev2018high,Liu:17}. Here we investigate the effect of the air slot on anapole excitations to realize high-efficiency metachains, when the substrate is vacuum in this part of simulations. Since the classic multipole expansion in spherical coordinates expands an arbitrary radiation pattern as an incoherent summation of orthogonal radiation patterns, it cannot independently investigate the effects of ED and TD modes. Therefore, we employ the Cartesian multipole decomposition method\cite{doi:10.1021/acs.nanolett.7b04200,
PhysRevB.94.205434} based on the EM fields obtained by finite-difference time-domain (FDTD) method (see Supplementary Material, eqs.(S1-S6) ) to identify the anapole states and discuss scattering contributions of multipole modes on both perfect disk (PD) and the split disk (SD). Figs. 1b-d) show the effect of multipole modes in PD and SDs with the air gap size $G$ = 0.5$R$ and 0.9$R$, respectively. An anapole mode requires the same value of ED moment ({\bf{p}}) and TD moment ($ik${\bf{T}}), and the out-of-phase condition of both moments. To be specific, the {\bf{p}} and $ik${\bf{T}} should satisfy arg({\bf{p}}) =  arg(-$ik${\bf{T}}). In this way, we can identify corresponding anapole frequencies, which are marked as $f_A$, $f_B$, and $f_C$ in Fig. 1. We can recognize the scattering dips of coherent ED and TD modes (solid orange lines) at these frequencies. Meanwhile, the electric field distributions that indicate the destructive interference of ED and TD modes are displayed in the insets of Figs. 1 b-d). Besides, we notice that the magnetic quadrupole (MQ) mode plays a more significant role with the increasing size of air gap $G$, which results in the total scattering efficiency at the anapole frequency varying from 0.9 to 6.4. The rising effect of MQ also eliminates the total scattering dip in SD with $G$ = 2.375$R$.

 To determine the near-field confinement capability of different disks, we calculate EM energy density respectively by $\int \varepsilon_{0} \varepsilon_{r}{\left | \mathbf{E} \right |}^2d V/V_p $ and $\int \mu_{0} \mu_{r}{\left | \mathbf{H} \right |}^2d V/V_p $, where $\varepsilon_{0}$ ($\mu_{0}$) is the permittivity (permeability) in vacuum and $\varepsilon_{r}$ ($\mu_{r}$) is the relative permittivity (permeability) when $V_p$ is the volume of particle. Fig. 1e) suggests that the split disk with $G$ = 0.5$R$ is the best candidate among these three particles for EM fields confinement due to the highest EM energy density at its anapole frequency $f_B$. On the other hand, the energy transfer of metachains depends not only on the energy confinement in an isolated particle but also on the near-field coupling efficiency. The near-field coupling strength can be estimated by comparing the localized electric energy inside the excited particle and its adjacent particle in a two-particle system\cite{mazzone2017near}. Fig. 1f) depicts that the coupling of anapoles happens at the near-field scale since the EM energy coupled in the adjacent particle sharply decreases with the increased separated distance. At the same time, the angular scattering distributions of particles in x-y plane ($z$=0) are given in the inset of Fig. 1f), which are collected by monitors located 3$R$ away from the particle center. In this inset, we find that the air gap structure can efficiently improve the scattering directionality of the anapole, which helps to enhance near field energy transfer. However, we should also note that the angular scattering distributions of the split disk with $G$ = 0.9$R$ have small scattering lobes along the y-axis, which suggests the radiative loss due to the MQ mode when $G$ is large.
 
\section*{Effect of periodicity on the metachain efficiency}

Since the two-particle system cannot entirely represent the behavior of metachains, we further investigate the one-dimensional chains of 21 periodically arranged particles, which is regarded sufficiently long for near-field energy transfer (see Fig.S1 in Supplementary Material). As illustrated in Fig. 2, the dielectric particle chains are excited by vertically incident Gaussian sources with E fields paralleled to $y$-axis at the center of the left-most particle. Figure 2a) exhibits the electric field distributions of metachains, which consist of arrays of PD, SD with $G$ = 0.5$R$, and SD with $G$ = 0.9$R$, with the periodicity $a$ = 2.15$R$. It is easy noticing that the SD chain with $G$ = 0.5$R$ has superior waveguiding efficiency. As noticed in Fig. 1, although the SD with $G$ =0.9$R$ has the highest coupling strength in the two-particle system, the long-chain efficiency of that is much lower because the radiative leaky MQ mode is dominant at $f_C$ frequency. Meanwhile, Fig. 2b) exhibits the corresponding electric field intensity evolution along the central axis of chains (x-direction) with periodicity $a$ = 2.15$R$ and 2.375$R$ . Due to the near-field decaying of anapole state, we can expect that the waveguiding efficiency at $f_A$ and $f_B$ decreases with the increasing periodicity, as shown in Figure 1f). However, the electric field guiding at $f_C$ frequency (Fig.2(b3)) gets slightly enhanced when the distance is larger than 10$a$, indicating a limited tuning effect from the periodicity. Moreover, we further discuss the effect of disorder in fabrication by randomly adding $x$-direction displacement $x_d$ for each particle in  Supplementary Material Fig S2, which suggests that the proposed metachain consisting of SD with $G$ = 0.5$R$ is insensitive to fabrication variations. 

Besides the high-efficiency transmission in a straight 1D chain, a low bending loss along the energy propagation is also a quest for subwavelength EM energy transfer in practical applications. The deviation from the straight path induces energy diffraction, leading to radiation losses and crosstalk to adjacent structures\cite{jiang2018low,xu2018ultra,wu2019ultra,wang2020ultra,fujisawa2017low,badri2020ultra}. Here, we discuss the loss at 90$^\circ$ bends of proposed metachains with different bending radiuses $R_B$ = 4$R$. The bending loss in a 90$^\circ$ bend is defined as the ratio that the beginning SD's electric energy to that of the end SD. For better comparison, we design the 90$^\circ$ bends working at optical communication band ($\sim $1550 nm) by scaling the designed metachains. Fig. 3 shows the electric field distributions of 90$^\circ$ bending metachains consisting of different particles with the bending radius $R_B = 4R$. For the SD metachain with $G$ = 0.5$R$, the designed $R$ = 0.57 \si{\um}, the bending radius $R_B$ = 2.26 \si{\um}, and the commensurate 90$^\circ$ bending loss is 0.32 dB. Compared with previous silicon-on-insulator (SOI) waveguides, shown in Supplementary Material, Table S1, the proposed metachain is ultra-thin and has a smaller bending radius when the bending loss is in an acceptable range. The bending radius of this metachain can be further reduced to 1.6 \si{\um} at the cost of increasing bending loss (0.81 dB). More details are given in Supplementary Material, Fig. S3. 

To reveal the physical mechanism of energy transfer in periodic metachains and the effect of periodicity $a$, we calculate the transmission and photonic band structures of metachains with $a$= 2.15$R$ using the FDTD method. As shown in Fig. 4, the energy transmission of metachains is shown by the colormap, indicating the energy transfer of metachains is closely related to the photonic bands. From the insets of Fig. 4, we confirm the excitations of anapole in the band structures by the electric field distributions, so we regard these bands anapole-induced. The light lines are marked by white dot-dashed lines. In Figs. 4 a-c), the anapole-induced bands of PD metachains are under the light line, when the anapole-induced bands of SD metachains cross with light lines, leading to different coupling effects. Specifically, the SD chains with $G$ = 0.5$R$ have a higher transmission at eigenmodes under the light line, when opposite behavior happens to the SD chains with $G$ = 0.9$R$.

To explain that fact, we introduce two energy ratios $ER_1$ and $ER_2$, which stand for the localized energy ratio of the whole SD structure including the air gap, and that of the silicon part of SD, respectively. From insets (b1-b3), we find that the $ER_1$ decreases when the eigenstate is above the light line due to the energy leaking. Meanwhile, the $ER_2$ gets enhanced when eigenstate approaches the light line in inset (b2) due to the coupling effects. In insets (c1-c2), we can also find that the $ER_1$ decreases while the eigenstate is above the light line. However, the $ER_2$ surprisingly increases when the eigenstate is leaky, which can be explained by the radiative MQ mode excited in SD with $G$ =0.9$R$. To be specific, eigenmodes below the light line localize more energy in the SD structure (including the air gap), leading to a higher radiative loss in the air gap due to the MQ mode in SD with $G$ = 0.9$R$, which is also suggested by the y-direction scattering lobes in Fig. 1f). Therefore, we find that $ER_1$ always decreases when coupling to the air, but $ER_2$ also depends on the scattering pattern of multipoles excited in isolated particles.

Therefore, we theoretically confirm that although the periodicity has a limited effect on the energy transfer by the coupling between light lines and anapole bands, the proposed metachains show a strong dependence on the excitations anapole states in individual particles. On the other hand, we notice that the proposed metachain can directly couple to the vertical incidence without couplers and transfer the energy with high-efficiency, when most of the existing subwavelength waveguides are either excited by in-plane incidence\cite{Cheng:14} or in need of couplers\cite{doi:10.1021/acs.nanolett.7b00381}. 

\section*{Near-field experimental analysis}

To verify the simulated results, we experimentally investigated the near-field signal of the proposed split disk metachain with $a$ = 2.375$R$, $G$ = 0.5$R$. The identical SDs in metachain are designed to have a radius $R$ = 3.85 \si{\um} to ensure that the metachain can be excited by the $\mathrm{CO_2}$ Laser (Access Laser L4G). We employed sapphire ($\mathrm{Al_2O_3}$) as the substrate, which has a relatively lower refractive index than silica at mid-infrared frequencies. The fabrication method of the metachain is given in Supplementary Material, Fig. S4, while the substrates' effect is discussed in Supplementary Material, Figs. S6 and S7. The scanning electron microscopy (SEM) image of the metachain is shown in Figs. 5a) and b), when the height of SD is measured by the atomic force microscopy (AFM), as shown in Fig. 5c). We notice that the measured geometry parameters slightly deviate from the designed parameters. The AFM-based s-SNOM (ANASYS NanoIR) system is shown in Fig. 5d), where the relative position between the laser spot and the platinum-coated tip remains unchanged during the scanning process. Thus, we can experimentally investigate the energy transfer performance by the ratio of signal under two situations: (i) the laser spot overlaps with the tip, which is the default setup of SNOM device; (ii) the in-plane distance between the laser spot and the tip is tuned to be about 37 \si{\um} along the metachain direction with the help of MCT detector and in-built optimizer. The collected near-field signal is directly related to the interplay between the EM energy confined in the metachain and the tip away from the laser spot. Since the in-plane tip-sample distance is fixed in one scanning process, the obtained signal has the same periodicity as the metachain. Therefore, we can investigate the performance by scanning a single particle inside the metachain. Multipole decomposition results of the fabricated particle are given in Supplementary Material, Fig S5, which show that the anapole mode appears around  10.36\si{\um}. Collected electrical near-field signal of particles in situations (i) and (ii) are respectively denoted as $\left | {Sg_{i}} \right | $ and $\left | {Sg_{ii}} \right | $. Consequently, the experimental in-plane signal ratio is defined as $\left | {Sg_{ii}/Sg_{i}} \right |$ , which are exhibited in Figure 5e) by blue solid squares. Note that the error bars denote the standard error of experimental results. 

Meanwhile, the numerical simulations on near-field signal acquisition are performed for comparison, when the effect of probe tip is considered. The platinum tip is modeled as a truncated cone capped a hemisphere at the bottom\cite{doi:10.1063/1.5008663,doi:10.1002/pssb.200983940}, which has a vertical tip-sample distance = 40 nm. Details of the numerical model are given in the Supplementary Material Fig. S9. Fig. 5e) shows that the in-plane signal ratio has its maximum at the anapole wavelength around 10.33 \si{\um}, and the experimental results are in good agreement with simulation results.

In addition, the excited modes in the metachain at different wavelengths can also be obtained by the s-SNOM system. Recorded raw near-field signals can be assumed as the linear superposition of $\mathbf{E_x}$, $\mathbf{E_y}$, and $\mathbf{E_z}$. Besides, due to the symmetry characteristic of y polarized incident laser and the periodicity, $\mathbf{ E_x}$, $\mathbf{ E_y}$, and $\mathbf{ E_z}$ distributions can be accordingly obtained\cite{doi:10.1021/acs.nanolett.7b04200} (see Supplementary Material Figs. S8). The results of SNOM Field decompositions and simulated electric fields at 10.17 \si{\um}, 10.33 \si{\um}, and 10.67 \si{\um} are exhibited in Fig. 6. Figures 6 a-c) show the amplitudes of $\mathbf{E_x}$, $\mathbf{E_y}$, $\mathbf{E_z}$, and the phase distributions of $\mathbf{E_z}$ when the metachain is excited by y-polarized source. In general, the experimental results agree well with the simulated results. In Fig. 6a), the wavelength is shorter than the anapole wavelength, $\mathbf{E_x}$ is relatively weak due to the dominant MQ response. In Fig. 6b), the $\mathbf{E_x}$ component gets enhanced and suggests the appearance of TD formed by the electric vortexes (as shown in Fig. 1c)) at the anapole wavelength. Meanwhile, the anapole state is indicated by the rapid change of $\mathbf{E_z}$ phase around the center of SD. It is also noticed at 10.33 \si{\um} that the $\mathbf{E_y}$ fields around the gap becomes more concentrated due to the interference of ED and TD. However, the anapole-induced Ey enhancement around the gap becomes ambiguous at 10.67 \si{\um} because of the dominant ED mode. Moreover, the near-field signals of metachain excited by x-polarized source at 10.33 \si{\um} are displayed in Fig. 6d). The electric fields are distinct from that of y-polarized cases since no anapole state is excited in that condition, which means the proposed SD particle chain is polarization-sensitive. In this way, the near-field signal of anapole state in the metachain is directly observed by the s-SNOM system, which is in accordance with the simulated results and confirms the vital role of anapole state on the subwavelength energy transfer.

\section*{Conclusions}

In conclusion, we have demonstrated ultracompact energy transfer in anapole-based metachains, which can confine EM energy to 1/13 of incident wavelength without couplers. The proposed metachain has high efficiency in both 1D and bending conditions for energy transfer, which is also insensitive to fabrication imperfections. We have performed the first near-field experimental measurement of energy transfer efficiency resulted from the coupling of excited anapoles, which have also been unambiguously identified by s-SNOM in this work. This insightful technique to transfer EM energy may find wide applications in realizing high-performance integrated photonic devices like a photonic network-on-chip\cite{ono2020ultrafast} or a quantum memory for entangled photons\cite{schwartz2016deterministic}, where chip-scale communications or routing will be needed. On the other hand, the anapole-based metachain provides a new platform for near-field thermal radiation transfer, which may lead to the radiative heat flux exceeding the blackbody limit in the far field by serval orders of magnitude\cite{kim2015radiative} and bring a revolution to the near-field radiative cooling for electronic devices.

\begin{acknowledgments}
The authors acknowledge the support from the National Natural Science Foundation of China (Grants No. 51636004 and No. 51906144), Shanghai Key Fundamental Research Grant (Grant No. 18JC1413300 and No. 20JC1414800), China Postdoctoral Science Foundation (Grants No. BX20180187 and No. 2019M651493), Open Fund of Key Laboratory of Thermal Management and Energy Utilization of Aircraft of Ministry of Industry and Information Technology (Grant No. CEPE2020015), and the Foundation for Innovative Research Groups of the National Natural Science Foundation of China (Grant No. 51521004).\end{acknowledgments}

\section*{Supplementary materials}
\noindent{Materials and Methods\\}
Supplementary Text\\
Figs. S1 to S10\\
Tables S1\\
References \textit{(16, 35, 39-44, 47, 48)}

\bibliography{scibib}

\clearpage

\begin{figure*}[p]
\centering\includegraphics[width=0.9\textwidth]{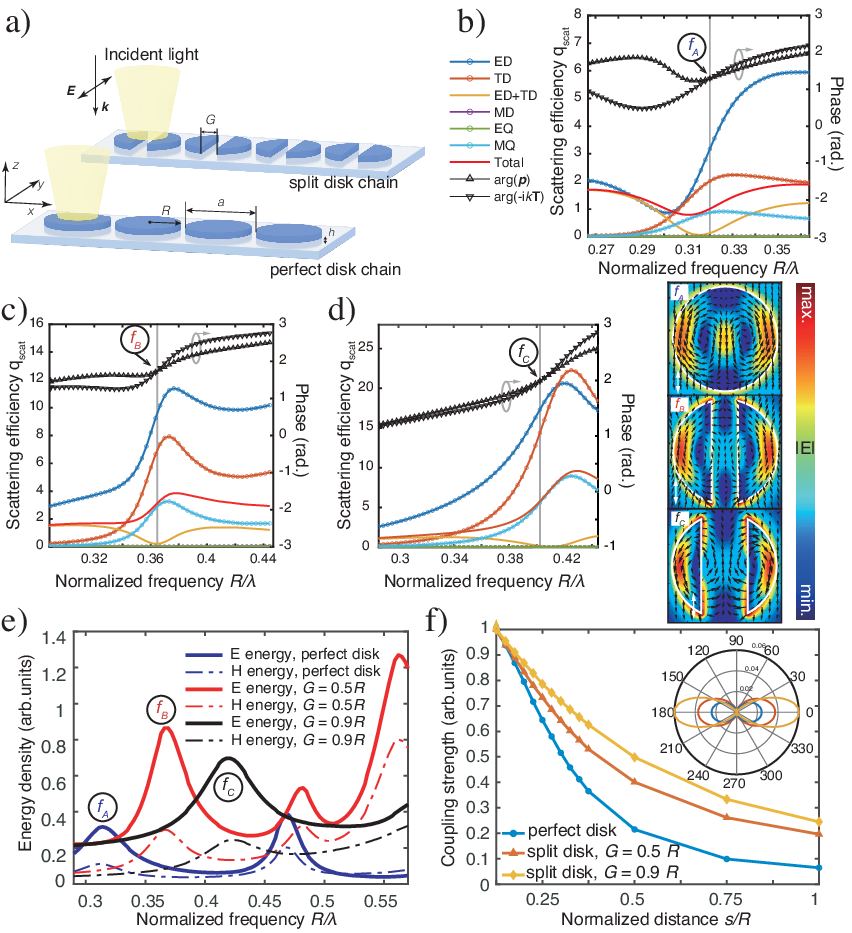}
\caption{\label{fig:1} Anapoles excited in isolated particles. a) Illustration of the dielectric subwavelength particles in proposed metachains. b-d) Multipole decomposition results of isolated perfect disks and split disks with $G$ =0.5$R$ and 0.9$R$, respectively. Insets show the electric field distributions of anapole states, while the solid black and white arrows represent the electric field vectors and the incident E field, respectively. e) The spectra of electric and magnetic energy density inside different particles. f) The evolution of the coupling strength due to the separation distance in the two-particle system, when angular scattering energy distributions in $x-y$ plane ($z$=0) of isolated particles are exhibited in the inset.}
\end{figure*}

\begin{figure*}[h]
\centering\includegraphics[width=0.6\textwidth]{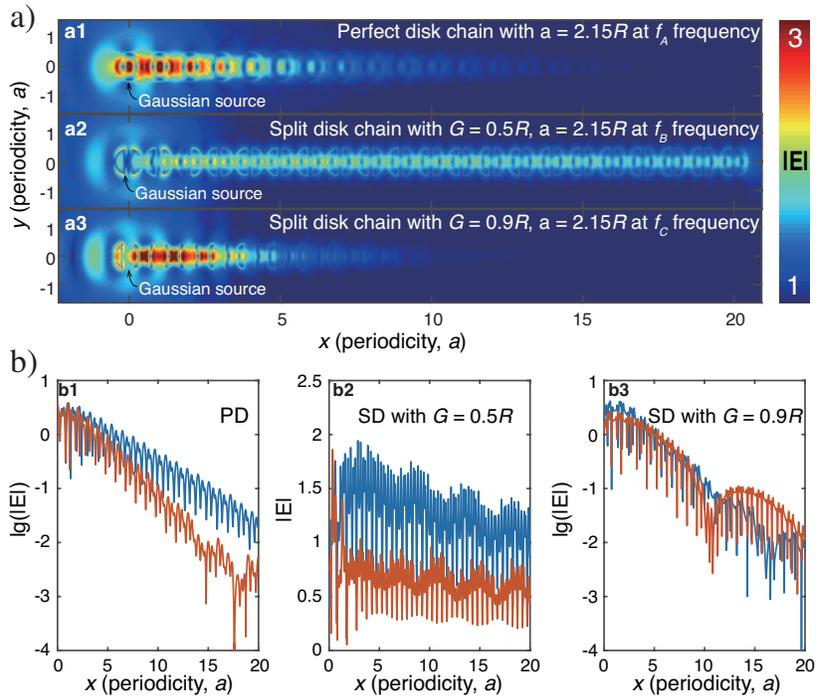}
\caption{\label{fig:2} a) The electric field distributions in metachains consisting of PD, SD with $G$ =0.5$R$ and 0.9$R$ at $f_A$, $f_B$, and $f_C$ frequencies with periodicity $a$ = 2.15$R$. b) The decaying of electric field intensity along the transmission direction (x-direction). The solid cyan lines denote the metachains with $a$ = 2.15$R$, while the solid red lines denote the metachains with $a$ = 2.375$R$.}
\end{figure*}

\begin{figure*}[h]
\centering\includegraphics[width=0.6\textwidth]{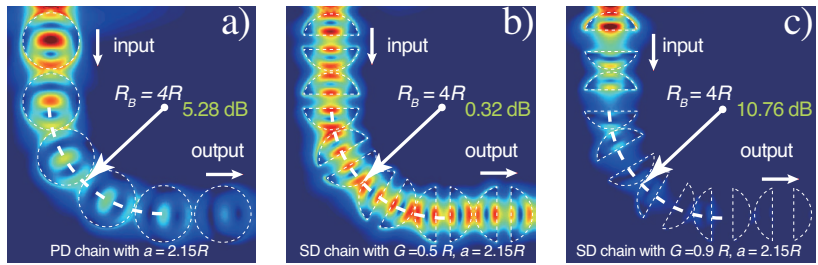}
\caption{\label{fig:3} The electric distributions of 90$^\circ$ metachains consisting of a) PD, b) SD with G = 0.5R, c) SD with $G$ = 0.9$R$, when bending radius $R_B$ = 4$R$. Corresponding bending losses are displayed in each subfigures.}
\end{figure*}

\begin{figure*}
\centering\includegraphics[width=0.6\textwidth]{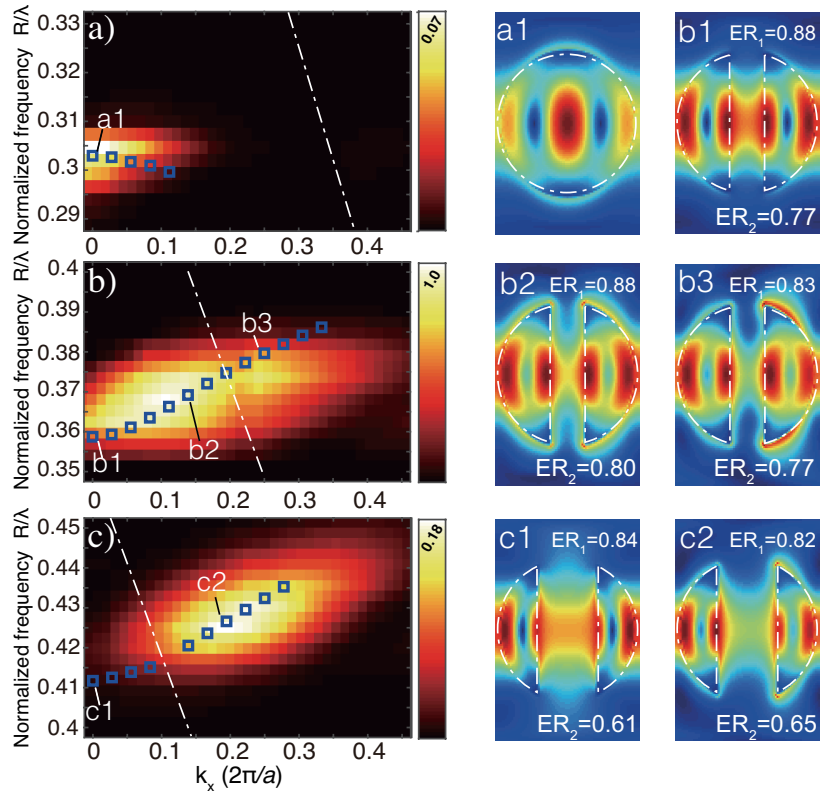}
\caption{\label{fig:4} Band structures and energy transmission of metachains. The band structures (cyan squares) and transmission (colormap) of metachains consisting of a) PD, b) SD with $G$ = 0.5$R$, c) SD with $G$ = 0.9$R$, when periodicity $a$ = 2.15$R$. Electric field distributions of specific points in bands are plotted in the insets.}
\end{figure*}

\begin{figure*}[h]
\centering\includegraphics[width=\textwidth]{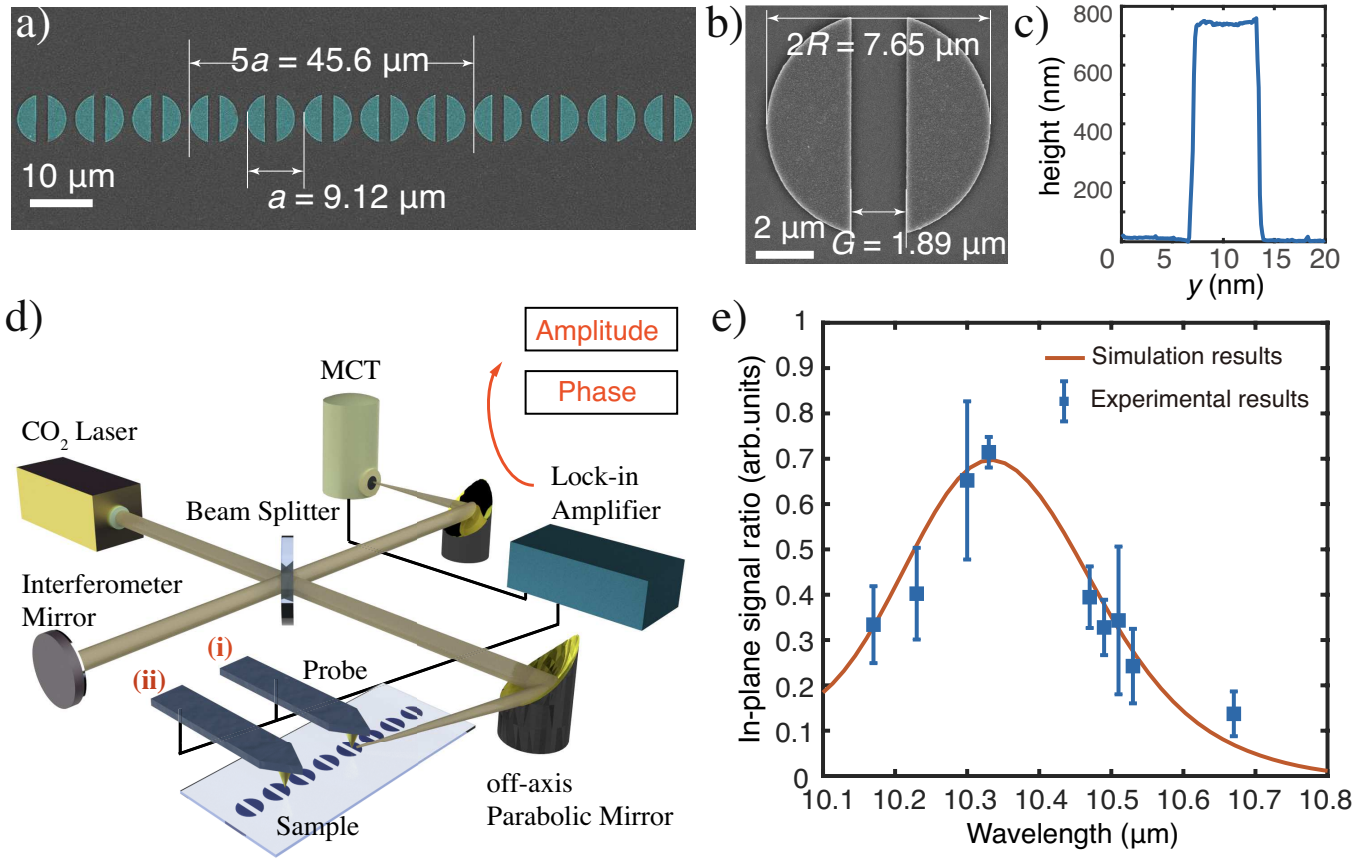}
\caption{\label{fig:6} Near-field experimental descriptions of the metachain. a) and b) show overall and enlarged SEM images of the fabricated metachain. c) AFM image of the fabricated SD height. d) Illustration of the s-SNOM system for near-field experimental verification of the metachain. e) The simulated and experimental results of in-plane near-field signal ratio, representing energy transfer performance. Note that the cyan error bars show the experimental standard error.}
\end{figure*}

\begin{figure*}[h]
\centering\includegraphics[width=\textwidth]{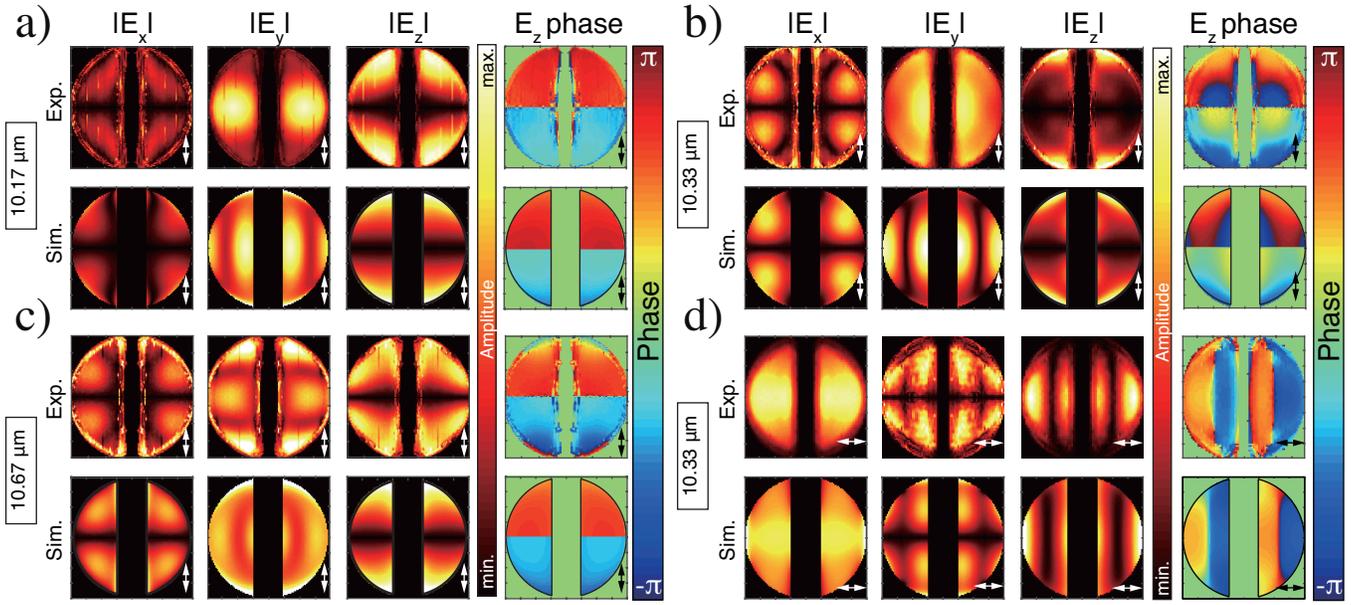}
\caption{\label{fig:7} Experimental and simulated electric field decompositions of the metachain. a-c), The fields decompositions of an isolated SD in the metachain excited by y-polarized Laser at a)10.17 \si{\um}, b) 10.33 \si{\um}, and c) 10.67 \si{\um}. d) The fields decompositions of an isolated SD in the metachain excited by x-polarized Laser at 10.33 \si{\um}. Note that the arrows denote the polarization of incident light.}
\end{figure*}
\end{document}